\documentclass[12pt]{article}
\usepackage{amssymb,exscale}
\usepackage{epsf}  

\def\be{\begin{equation}}
\def\ee{\end{equation}}
\def\bea{\begin{eqnarray}}
\def\eea{\end{eqnarray}} 
\def\nn{\nonumber \\}

\def\tfrac#1#2{{\textstyle{#1\over #2}}}
\def\half{\tfrac{1}{2}}
\def\x{\times}

\def\part{\partial}

\def\bpart{{\bar{\partial}}}
\def\tnabla{{\tilde\nabla}}
\def\bnabla{{\bar\nabla}}
\def\hnabla{{\hat\nabla}}

\def\hC{{\hat C}}
\def\hF{{\hat F}}

\def\hR{{\hat R}}
\def\hT{{\hat T}}

\def\tD{{\tilde D}}
\def\tR{{\tilde R}}

\def\bR{{\bar R}}

\def\hg{{\hat g}}

\def\hs{{\hat s}}

\def\tg{{\tilde g}}

\def\bh{{\bar h}}

\def\hphi{{\hat{\phi}}}

\def\tchi{{\tilde{\chi}}}

\def\hGamma{{\hat{\Gamma}}}

\def\makeatletter{\catcode`\@=11}
\makeatletter
\def\mathbox#1{\hbox{$\m@th#1$}}%
\def\math@ccstyles#1#2#3#4#5#6#7{{\leavevmode
      \setbox0\mathbox{#6#7}%
      \setbox2\mathbox{#4#5}%
      \dimen@ #3%
      \baselineskip\z@\lineskiplimit#1\lineskip\z@
      \vbox{\ialign{##\crcr
             \hfil \kern #2\box2 \hfil\crcr
             \noalign{\kern\dimen@}%
             \hfil\box0\hfil\crcr}}}}
\def\mathaccstyles{\math@ccstyles\maxdimen}
\def\maththroughstyles{\math@ccstyles{-\maxdimen}}
\def\unitmatrixDT%
 {\maththroughstyles{.45\ht0}\z@\displaystyle {\mathchar"006C}\displaystyle 1}
\begin{document}

\rightline{IFT-UAM/CSIC-99-38}
\rightline{hep-th/9910077}
\rightline{\today}
\vspace{1truecm}

\centerline{\Large \bf Curved branes and cosmological $(a, b)$-models}
\vspace{1truecm}

\centerline{
    {\bf Bert Janssen}\footnote{E-mail address: 
                                  {\tt bert.janssen@uam.es}}
                                                                }
\vspace{.4truecm}
\centerline{{\it Instituto de F{\'\i}sica Te{\'o}rica, C-XVI,}}
\centerline{{\it Universidad Aut{\'o}noma de Madrid}}
\centerline{{\it E-28006 Madrid, Spain}}
\vspace{2truecm}

\centerline{\bf Abstract}
\vspace{.5truecm}

\noindent
We construct $p$-brane solutions with non-trivial world volume metrics 
and show that applied to supergravity theories, they will
lead to threshold BPS bound states of intersecting solutions. However
applied to certain specific values of the couplings in cosmological 
$(a,b)$-models non-trivial solutions can be constructed.

\newpage

{\bf 1. Introduction} 
\vspace{.3cm}

\noindent
It is very well known that supergravity theories have solutions in the form of
extended object, the so-called $p$-branes \cite{Stel}. They appear as 
solitonic objects that couple to the metric $\hg_{\mu\nu}$, dilaton $\hphi$ and
$(p+1)$-form gauge fields $\hC$ of the theory \cite{Hull}. In general they are 
of the following form:
\bea
d\hs^2&=& H^\alpha \eta_{mn} dx^mdx^n - H^\beta \delta_{ij}dy^idy^j \ , \nn
e^{-2\hphi}&=& H^\gamma\ , 
\label{flatbrane} \\
\hC_{01...p}&=& \delta \ H^{-1}  \ ,      \nonumber
\eea
where the parameter $\alpha, \beta, \gamma$ and $\delta$ are determined by the 
type of $p$-brane and the specific supergravity theory considered. The 
coordinates $x^m$ are the world volume coordinates and $y^i$ the coordinates 
transverse to the brane. The above solution generally breaks half of the 
spacetime supersymmetry.

\noindent
There have been several attempts to generalise the above solution by trying to 
write down more general metrics $\bh_{ij}$ over the  transverse space than the 
conformally flat metric shown above \cite{DLPS, GGPT}. It was demonstrated 
that this can be done as long as the transverse space is Ricci-flat and that
part of the supersymmetry is preserved in the presence of Killing spinors. 
On the other hand, in \cite{Empar} a solution was presented of the near-horizon
limit of an M2-brane wrapped over an  arbitrary genus surface.

\noindent
Recently there has been a attempt \cite{Per} to generalize the standard 
$D8$-brane solution of \cite{BRGPT} by imposing a general metric $\tg_{mn}$ 
on the world volume of the domain wall. Also here it was found that the 
equations of motion and supersymmetry were satisfied, provided that the 
metric $\tg_{mn}$ is Ricci-flat and allows covariantly constant Killing 
spinors.\footnote{Shortly after this paper, another paper \cite{FoF} on 
Ricci-flat branes appeared.} It was shown that such solutions can be 
constructed, choosing as world volume a manifold with the appropriate holonomy 
groups.

\noindent
Here we want to take a different way and try to look for solutions to the 
Einstein-dilaton action coupled to gauge fields. We will show that for 
standard supergravity 
theories, i.e. in the absence of a cosmological constant, these solutions 
exist, but can be seen as solutions made out of 
inter\-sec\-ting branes. However, curved $p$-branes can be used to construct
solutions to general Einstein--dilaton-gauge field theory with a cosmological 
constant, the so-called cosmological $(a,b)$-models, for specific values of 
the couplings $a$ and $b$. These theories in general are 
not supersymmetric. Nevertheless, it is interesting to see how curved branes 
fit into this setup to provide a new class of solutions.      

\noindent
The organisation of this letter is as follows: 
in section 2 we will set our notation  by giving a general Ansatz for curved 
$p$-brane and compute the various curvature tensors of our Ansatz. 
In section 3, we will look at the equations of motion of the curved $p$-branes 
in a general supergravity theory, reproduce the results of 
\cite{DLPS, GGPT, Per} and give solutions satisfying the conditions.
In section 4 we will apply the Ansatz to construct solutions to cosmological 
$(a,b)$-models.

\vspace{.3cm}
{\bf 2. The curved $p$-brane Ansatz}
\vspace{.3cm}

\noindent
We start with the following Ansatz for the curved $p$-brane metric in $D$ 
dimensions:
\be
d\hs^2 = H^\alpha(y) \tg_{mn}(x) dx^mdx^n 
                    - H^\beta(y) \bh(y)_{ij} dy^i dy^j. 
\label{Ansatz}
\ee
We allow the internal metric $\tg_{mn}(x^m)$ over the world volume to depend 
on the world volume coordinates $x^m$ and the transverse space to 
have a more general geometry, described by the external metric $\bh_{ij} (y)$. 
The parameters $\alpha$ and $\beta$ are to be choosen such that in the case of 
flat internal and external metrics ($\tg_{mn}=\eta_{mn}$ and 
$\bh_{ij}= \delta_{ij}$), the standard (``flat'') $p$-brane solutions 
(\ref{flatbrane}) are recovered. In 
that case the function $H(y)$ is harmonic in the transverse coordinates: 
$\part_i\part_i H =0$.    

\noindent
It is straight forward to compute the Ricci tensor and the Ricci scalar for the
metric (\ref{Ansatz}):
\bea
\hR_{mn} &=& \tR_{mn} \ -  \
             \tfrac{\alpha}{2} \tg_{mn} H^{\alpha - \beta -1} \ 
               \Bigl [ \bnabla_i \bpart^i H 
                        + (\tD - \beta -1) (\bpart H)^2 \Bigr ] \ ,   \nn
\hR_{im} &=& 0 \ ,          \nn
\hR_{ij} &=& \bR_{ij} \ 
             +\ \tfrac{\beta}{2} H^{-1} \bnabla_k \bpart^k H\ \bh_{ij}
             + (\tD - \beta) H^{-1} \bnabla_i \part_j H   \nn
         &&  
          + \ \tfrac{\beta}{2} (\tD- \beta-1) H^{-2} (\bpart H)^2 \ \bh_{ij} \\
         &&    
          + \Bigl[ \tfrac{\alpha^2}{4} (p+1) +  \tfrac{\beta^2}{4} (D-p-1) 
                    - \tD(\beta + 1) + \tfrac{\beta^2}{2} + \beta  \Bigr]
                   H^{-2} \part_i H \part_j H \ , \nn
\hR &=&  H^{-\alpha} \tR \ - \ H^{-\beta} \bR   
         \ -\  (2\tD - \beta) H^{-\beta-1} \bnabla_i \bpart^i H  \nn
     &&   \hspace{-.4cm}
     -  \Bigl[ \tfrac{\alpha^2}{4} (p+1) +  \tfrac{\beta^2}{4} (D-p-1) 
                    + \tD(\tD -2\beta -2) + \tfrac{\beta^2}{2} + \beta  \Bigr]
                         H^{-\beta-2} (\bpart H)^2.\nonumber
\label{tensors}
\eea
Here $\tR_{mn}\ (\bR_{ij})$ and $\tR \ (\bR)$ are the Ricci tensor and
the Ricci scalar of the internal (external) metric, $p$ is the number of 
spatial dimensions of the brane and 
\be
\tD= \tfrac{\alpha}{2}(p+1) + \tfrac{\beta}{2} (D-p-1).
\ee   
We denote by $\bnabla_i$ the covariant derivative with respect to the metric 
$\bh_{ij}$. Furthermore we have defined
\be
\bpart^i H = \bh^{ij}\part_j H,
\hspace{1cm}
(\bpart H)^2 = \bh^{ij} \part_i H \part_j H \ .
\ee 
Note that the overall Ricci tensor and Ricci scalar (\ref{tensors}) 
factorize into parts coming from the internal and external metric and a part 
coming from the overall metric of the brane.


\vspace{.3cm}
{\bf 3. Ricci-flat branes and intersections}
\vspace{.3cm}

\noindent
We will now look at the concrete example of a $p$-brane in supergravity. Our 
Ansatz for the curved $p$-brane solution is given by:
\bea
d\hs^2 &=&  H^{\alpha} \Bigl[\tg_{mn} dx^m dx^n \Bigr]
               - H^{\beta} \Bigl[ \bh_{ij} dy^i dy^j \Bigr] \ , \nn
e^{-2\hphi} &=& H^{\gamma} \ , \label{D3} \\
\hC_{01..p}&=& \sqrt{|\tg|} \ H^{-1} \ . \nonumber
\eea
Note that the gauge fields are all given in the electric formulation, where the dual 
potentials are used for magneticly charged $p$-branes $(p>3)$.
The extra factor $\sqrt{|\tg|}$  has been added to 
compensate for the curvature of the internal metric in the equations of motion.
The equation of motion of $\hC$ is satisfied if the function $H(y)$ is 
harmonic on the external metric $\bh_{ij}$:
\be
\bnabla_i \bpart^i H = 0 \ .
\label{harm}
\ee
The Einstein and dilaton equations factorize completely into a part coming 
from the internal and external metrics and a part that is identical zero, 
being the equations of motion of the ``flat'' $p$-brane, upto 
Eqn.~(\ref{harm}). Therefore we have that:
\bea
&& \tR_{mn} = 0 \ ,   \nn
&& \bR_{ij}=0 \ , \label{tensoreqn}\\
&& H^{-\alpha} \tR \  - \ H^{-\beta} \bR \ =\  0 \  . \nonumber
\eea   
These are the Ricci-flatness conditions given in \cite{DLPS, GGPT, Per}. Also 
the supersymmetry variations factorize in a part proportional to the projection
operator of the $p$-brane and the following conditions on $\epsilon$ 
coming from the internal and external metrics:
\be
\tnabla_m \epsilon = 0 \ ,   \hspace{1.5cm}
\bnabla_i \epsilon = 0 \ .
\label{susy}
\ee 
The problem now consists of finding (non-trivial) $(p+1)$- or 
$(D-p-1)$-dimensional manifolds admitting Killing spinors. These are given in
terms of their holonomy groups: only those manifolds are allowed that have 
holonomy groups Spin(7), $G_2$, $SU(3)$ and $Sp(1)$ (or subgroups thereof), 
depending on the dimension of the manifold considered \cite{Per, FoF}.

\noindent
Here however we want to take a different way and try to satisfy the conditions 
(\ref{tensoreqn})-(\ref{susy}) by looking at Ricci-flat supersymmetric 
solutions of the $(p+1)$- or $(D-p-1)$-dimensional Einstein-dilaton-gauge 
field Lagrangian. To our knowlegde, there are two such solutions: the 
Kaluza-Klein monopole \cite{GP} and the gravitational wave \cite{Brinkm}.
 
\noindent
Choosing for the internal metric $\tg_{mn}$ the four-dimensional gravitational 
wave, the generalized $p$-brane (\ref{D3}) takes the form:
\bea
&& 
 d\hs^2 = H^{\alpha}\Bigl[(2-F)dt^2 - Fdz^2 - 2(1-F)dtdz - dx_a^2 \Bigr]
  -   H^{\beta} d\vec{y}^2 \nn
&& e^{-2\hphi}= H^\gamma \ , \hspace{4cm} \\
&& \hC_{01...p}=  H^{-1} \ .  \nonumber
\label{D3W}
\eea
This solution satisfies the equations of motion if the functions $F$ and $H$
are harmonic in $x^a$ and $y^i$ respectively and preserves one quarter of 
supersymmetry.
However it is not difficult to see that this is actually the solution of a 
theshold BPS intersection of a $p$-brane and a ten-dimensional wave in its 
world volume \cite{Tseyt, RT} (in the notation of \cite{BREJS}):
\bea
\begin{array}{rcc}
(1| p,{\cal W})=
      & \multicolumn{2}{l}{\left\{ {\begin{array}{c|ccccccccccc}
                                   \x&\x&\x&\x&\x&\x&-&-&-&-& &:H  \\
                                   \x&z& -& -& -& -&-&-&-&-& &\ \,\,:F \ .\\
                                 \end{array}} \right.                  }   \\
      &\hspace{1cm}
       \hbox to 0pt{\hss$\underbrace{\hskip1.0cm}_{t,z}$\hskip 3mm\hss}
          \hspace{2cm}
          \hbox to 0pt{\hss$\underbrace{\hskip2.0cm}_{x_a}$\hskip 3mm\hss}
         &\hspace{-0.9cm}
         \hbox to 0pt{\hss$\underbrace{\hskip1.0cm \hskip 1.2cm}_{y_i}$\hss} 
\end{array}
\label{D3Wi}
\eea
Also the inclusion of a Kaluza-Klein monopole in the transverse space will lead
to a known threshold BPS intersection of a $p$-brane and a monopole
\cite{Tseyt, RT, BREJS}
\be
 (p|p, {\cal KK})=
\mbox{ {\scriptsize
$\left\{ \begin{array}{c|ccccccccc}
         \x & \x & \x & \x  & -  & -  & -  & - & -  & -   \\           
         \x & \x & \x  & \x  & \x & \x &  z & A_1 & A_2 & A_3   
                                         \end{array} \right.$}}  \ .
\label{D3KK}  
\ee
The dependence of the harmonic functions $H$ and $F$ (in the language of 
intersecting branes, on the overall transverse and relative transverse 
coordinates respectively, i.e. the gravitational wave delocalised in its 
overall transverse directions) is the correct one to belong to the class of 
threshold BPS bound states \cite{BBJ}.

\noindent
A logical next step would be trying to weaken the Ricci-flatness condition by 
exciting other fields with dependences on the world volume coordinates (For 
example excite a $U(1)$ gauge field and construct a extreme 
Reissner-N\"ordstr\"om solution on the world volume). 
However, these generalisations turn out to be impossible: adding extra 
gauge fields or dilaton with dependence on the world volume coordinates 
inevitably leads to terms in the equations of motion which cannot be cancelled.
For example adding a dilaton $\hphi \sim \tchi(x)$ will lead in the Einstein 
equations to a term
\be
\hnabla_i \part_m \hphi= \hGamma_{im}^n \part_n \tchi  
                       = \frac{1}{4} H^{-1} \part_i H \ \part_m \tchi , 
\ee
which clearly cannot be cancelled by other terms. 
It would be interesting to see whether the Ricci-flatness condition can be 
circumvented by coupling a curved $p$-brane to a world volume action and let 
the Born-Infeld field on the $D$-brane play the role of $U(1)$ gauge field of
an extreme Reissner-N\"ordstr\"om solution.


\vspace{.3cm}
{\bf 4. Curved branes and cosmological $(a,b)$-models}
\vspace{.3cm}

\noindent
As was pointed out in the previous section, the Ricci-flatness condition is 
restriction, leading either to the class of branes with world volumes manifolds
with the adequate holonomy groups, or to threshold BPS bound state
intesections of $p$-branes with gravitational waves or monopoles. 
 
\noindent
However, if one forgets for one moment the Ricci-flatness condition imposed by 
the first two equations of (\ref{tensoreqn}), the dilaton equation admits a 
more general solution:
\be
\tR = \kappa  \ , 
\hspace{2cm}
\bR= \kappa \ H^{\beta-\alpha}
\ee
where $\kappa$ is an arbitrary constant. The obvious way now to make this 
solution also fit the Einstein equations of (\ref{tensoreqn}), is to modify 
these equations by introducing a cosmological constant in the theory.

\noindent
So we can  use curved branes to obtain solutions of theories with 
cosmological constants. The idea consists of taking a $p$-brane solution of a 
theory with zero cosmological constant, impose on the world volume of the 
brane a metric with (non-zero) constant curvature and see whether the new 
Ansatz satisfies the equations of motion of the theory with non-zero 
cosmological constant. In other words, the question is whether we can relate 
the cosmological constant of the world volume metric to the cosmological 
constant of the full theory.   

\noindent
The most general theory in $D$ dimensions of gravity coupled to a dilaton and 
a $(p+1)$-form gauge field in the presence of a cosmological constant is the 
so-called cosmological $(a,b)$-model:
\be
{\cal L}_D = \sqrt{|\hat g|} \Bigl\{ e^{-2\hphi} 
                \Bigl[\hat R - 4(\part\hphi)^2 \Bigr]
             +\tfrac{(-)^{p+1}}{2(p+2)!}\ e^{a\hphi}\ \hF_{(p+2)}^2
             + e^{b\hphi}\Lambda \Bigr\} \ .
\ee
The parameters $a, b$ are constants parametrising the dilaton potential that
couples to the gauge fields and the cosmological constant. Note that the above 
Lagrangian will only be supersymmetric for specific values of  $a$ and $b$. 
For example, for $D=10,\ a=b=0,$ we recover Romans' massive supergravity 
theory \cite{Rom}, and for $D< 10, \ a=b=-2$  massive heterotic supergravity 
\cite{BRE, MS}.

\noindent 
The equations of motion for this model are given by:
\bea
&&\hR_{\mu\nu}  - \tfrac{b+2}{4} e^{(b+2)\hphi} \Lambda \hg_{\mu\nu} 
                - 2\hat\nabla_\mu \part_\nu\hphi       
                - \tfrac{1}{(p+1)!} e^{(a+2)\hphi} \ \hT_{\mu\nu} (\hF)= 0, \nn
&&\hR + 4(\part\hphi)^2 - 4 \hat\nabla^2 \hphi 
      -\tfrac{a}{4(p+1)!} e^{(a+2)\phi}\hF^2 
      -\tfrac{b}{2} e^{(b+2)\phi} \Lambda =0,       \label{EOM}\\
&& \hat\nabla_\mu \Bigl[ e^{a\phi} \hF^{\mu\rho_1 ...\rho_{p+1}} \Bigr]=0,
                        \nonumber
\eea
where $\hT_{\mu\nu} (\hF)$ is the energy-momentum tensor of the gauge field
\be
\hT_{\mu\nu} (\hF) = 
      \hF_{\mu\rho_1 ...\rho_{p+1}}\hF_\nu{}^{\rho_1 ...\rho_{p+1}}
                      - \tfrac{1}{2(p+2)}\hg_{\mu\nu} \hF^2 .  
\ee
A general $p$-brane solution in $D>2$ for the case of $\Lambda=0$ is of the 
form \cite{LPSS, Berg}:
\bea
d\hs^2 &=& H^\alpha dx_m^2 - H^\beta dy_i^2 \ , \nn
e^{-2\hphi} &=& H^\gamma \ ,  \label{stellebrane}\\
\hF_{01..pi} &=& \delta\  \part_i H^{-1} \ , \nonumber
\eea    
with the parameters $\alpha,\beta, \gamma$ and $\delta$ taking the values
\bea
\alpha &=&   \tfrac{2-a}{N}  \ , \hspace{1.5cm}
\beta   =  - \tfrac{2+a}{N}  \ , \hspace{1.5cm}
\delta^2  =  - \tfrac{4}{N}    \ , \nn
\gamma &=& - \tfrac{1}{N} \Bigl[ 2(p+1) - \half (a+2)(D-2) \Bigr] \ , 
\label{cond} \\
N  &=& (p+1)a - \half (D-2)(1+\tfrac{a}{2})^2 \ . \nonumber
\eea
The aim is now to put a non-trivial world volume metric on these brane 
solutions and see whether we can construct in this way solutions to the full 
equations of motion (\ref{EOM}). These solutions are of course not the most 
general solutions to these equations. On the contrary, we are looking at a
very specific class of solutions to cosmological $(a,b)$-models, involving 
curved $p$-branes. Other $p$-brane solutions to various cosmological 
theories where given in \cite{MS1, LMPX, Singh, MS, JMO}.  

\noindent
Plugging Ansatz (\ref{D3}) into the equations (\ref{EOM}), we obtain,
upto the equations of motion of the ``flat'' $p$-brane:
\bea
&&\hR - \tfrac{b}{2} e^{(b+2)\hphi} \Lambda \ = \
          H^{-\alpha} \tR \ - \  H^{- \beta} \bR 
          \ - \  \tfrac{b}{2} H^{- \tfrac{b+2}{2}\gamma}\Lambda 
                                             \ \equiv \  0 \ ,\nn
&&\hR_{mn} - \tfrac{b+2}{4} e^{(b+2)\hphi} \Lambda \ \hg_{mn} \ = \
 \tR_{mn}\ -\ \tfrac{b+2}{4}\Lambda \ H^{\alpha-\tfrac{b+2}{2}\gamma}\ \tg_{mn}
                                             \ \equiv \ 0  \ , 
\label{cosmvgl} \\
&&\hR_{ij} - \tfrac{b+2}{4} e^{(b+2)\hphi} \Lambda \ \hg_{ij} =
 \bR_{ij}\ +\ \tfrac{b+2}{4} \Lambda\ H^{\beta -\tfrac{b+2}{2}\gamma}\ \bh_{ij}
                                             \ \equiv \ 0  \ . \nonumber
\eea 
These equations can be satisfied if
\bea
\tR_{mn} &=& -\tfrac{1}{D-2}\Lambda \ \tg_{mn} \ ,  \hspace{1cm}
\bR_{ij} = \tfrac{1}{D-2}\Lambda\ H^{\beta + \tfrac{2\gamma}{D-2}}\  \bh_{ij} \ , \nn
\gamma  &=& - \tfrac{D-2}{2} \alpha  \ ,\hspace{2cm}
b =- \tfrac{2D}{D-2} \ , \hspace{1cm} \label{cosmcond} 
\eea
Note that the parameter $b$ only depends on the dimension of the overall space:
for every dimension there is a well-defined cosmological dilaton potential 
for which curved cosmological $p$-branes are allowed.
The condition for $\tg_{mn}$ is the expression for the Ricci tensor of a 
constant curvature space, for example a  
$(p+1)$-dimensional (anti)-de Sitter space, with cosmological constant 
$\tfrac{\Lambda}{D-2}$ (depending on the sign of $\Lambda$).
The condition for the metric $\bh_{ij}$, together with the harmonicity 
condition (\ref{harm}), is much more involved and it is not 
clear what the solution is to these equations. 

\noindent
Combining the conditions (\ref{cond}) for the ``flat'' $p$-brane solution with 
the above conditions (\ref{cosmcond}), we find that we have cosmological 
solutions for curved $p$-branes with 
\be
p = D-3\ .
\label{p-eqn}
\ee 

\noindent
Let us now look at some explicite examples: 

\begin{itemize}
\item {\bf Romans' theory:} It is easy to see that there are no solutions of 
the above class in ten-dimensional massive Type IIA supergravity, since the 
condition (\ref{cosmcond}) on $b$ is not compatible with the coupling of the 
cosmological constant in Romans' theory ($b=0$). Note that the curved D8-brane 
solutions of \cite{Per} is not found in this setup. This is because the curved 
D8-brane falls outside our Ansatz: the ``flat'' D8-brane is not a solution of 
the type given in (\ref{stellebrane})-(\ref{cond}).

\item {\bf Massive heterotic supergravity:} From Eqn.~(\ref{cosmcond}) it 
follows that $b$ can not take the value $-2$ in this setup. Therefore there are
no curved cosmological $p$-branes in massive heterotic supergravity. 

\item {\bf $p$-branes in $D=10$:}  From Eqn. (\ref{p-eqn}), we see that the 
Type IIB D7-brane coupled to RR 9-form $(a=0)$ can be embedded in a 
cosmolo\-gical theory, with dilaton coupling for the cosmological term 
$b=-5/2$,
the only coupling allowed in ten dimensions in this setup. Of course, this
cosmological theory is no longer a supergravity theory. 
If we choose for the constant curvature surface on the world volume an 
eight-dimensional AdS space in horospheric coordinates, the curved 
cosmological D7-brane is of the form
\bea
&&d\hs^2= H^{-1/2} 
        \Bigl[ \tfrac{56}{\Lambda x_7^2}( dt^2- dx_1^2 - ... -dx_7^2 ) \Bigr] 
                     - H^{1/2} \bh_{ij}dy^idy^i\ , \nn  
&&e^{-2\hphi} \ = \  H^2 \ ,\\
&&\hF_{01..7i}\ = \ \part_i H^{-1} \ ,\nonumber
\eea  
where the metric $\bh_{ij}$ has to satisfy 
$\bR_{ij}= \tfrac{\Lambda}{8} H \ \bh_{ij}$.

\item {\bf $p$-branes in $D=6$:} In six dimensions, we can embedd the
curved D3-brane in a cosmological theory with dilaton coupling $b=-3$:
\bea
&&d\hs^2= H^{-1} 
        \Bigl[ \tfrac{12}{\Lambda x_3^2} (dt^2- dx_1^2 - ... -dx_3^2 ) \Bigr] 
                     - H \ \bh_{ij} dy^idy^i\ , \nn 
&&e^{-2\hphi} \ = \  H^2 \ ,\\
&&\hF_{01..7i}\ = \sqrt{2} \ \part_i H^{-1} \ . \nonumber
\eea

\item {\bf $p$-branes in $D=4$:} In four dimensions, strings can be embedded 
in a theory with dilaton coupling $b=-4$. 
   
\item $\beta = - \tfrac{2\gamma}{D-2}$: For this value condition 
(\ref{cosmcond}) on $\bR_{ij}$ would become very simple. However, as can be 
seen from Eqns. (\ref{cond}), the constraint $\alpha= \beta$ leads 
to inconsistencies.   
\end{itemize}

\vspace{.3cm}
{\bf 5. Conclusions}
\vspace{.3cm}

\noindent
We showed that putting a Ricci-flat metric in the world volume of a $p$-brane 
in supergravity leads to BPS threshold intersections of the considered 
$p$-brane with a gravitational wave or a Kaluza-Klein monopole. Going beyond 
the Ricci-flatness condition we found a new class of curved $p$-branes 
solutions in theories with a particular coupling of the cosmological constant. 

\vspace{.5cm}

{\bf Acknowledgments}

\noindent
The author wants to thank C\'esar G\'omez, Tom\'as Ort{\'\i}n and especially 
Patrick Meessen for the usefull discussions and valuable ideas.
The work of B.J. has been supported by the TMR program FMRX-CT96-0012 on 
{\sl Integrability, non-perturbative effects, and symmetry in quantum field 
theory}.


\begin{thebibliography}{99}
\bibitem{Stel} For  a review, see e.g.: K. S. Stelle,
               {\it BPS Branes in Supergravity}, hep-th/9803116
               and references therein

\bibitem{Hull} C. M. Hull, Nucl.Phys. B509 (1998) 216, hep-th/9705162

\bibitem{DLPS} M. J. Duff, H. L\"u, C. N. Pope and E. Sezgin,
               Phys.Lett. B371 (1996) 206, hep-th/9511162  

\bibitem{GGPT} J. P. Gauntlett, G. W. Gibbons, G. Papadopoulos and 
               P. K. Townsend,
               Nucl.Phys. B500 (1997) 133, hep-th/9702202

\bibitem{Empar} R. Emparan,  Phys.Lett. B432 (1998) 74, hep-th/9804031

\bibitem{Per} D. Brechner and M.J. Perry, {\it Ricci-flat branes}, 
              hep-th/9908018

\bibitem{BRGPT} E. Bergshoeff, M. de Roo, M. Green, G. Papadopoulos and 
                P.Townsend, Nucl.Phys. B470 (1996) 113, hep-th/9601150

\bibitem{FoF} J.M. Figueroa-O'Farrill, {\it More Ricci-flat branes},
              hep-th/9910086 

\bibitem{GP} D. J. Gross and M.J. Perry, Nucl.Phys. B226 (1983) 29; \\
             R. D. Sorkin, Phys. Rev. Let. 51 (1983) 87

\bibitem{Brinkm} H.W. Brinkmann, Proc. Nat. Acad. Sci. 9 (1923) 1

\bibitem{Tseyt} A. A. Tseytlin, Nucl.Phys. B475 (1996) 149, hep-th/9604035

\bibitem{RT}  J.G. Russo and A.A. Tseytlin, Nucl.Phys. B490 (1997) 121,
              hep-th/9611047

\bibitem{BREJS} E. Bergshoeff, M. de Roo, E. Eyras, B. Janssen and J.P. van der
                Schaar, Class. Quant. Grav. 14 (1997) 2757, hep-th/9704120  

\bibitem{BBJ} K. Behrndt, E. Bergshoeff and B. Janssen
              Phys. Rev. D55 (1997) 3785, hep-th/9604168 

\bibitem{Rom} L.J. Romans, Phys. Lett. B (1986) 374

\bibitem{BRE} E. Bergshoeff, M. de Roo and  E. Eyras,
               Phys.Lett. B413 (1997) 70, hep-th/9707130

\bibitem{MS} J. Maharana and H. Singh, Phys.Lett. B408 (1997) 164, 
             hep-th/9705058

\bibitem{LPSS} H. L\"u, C. N. Pope, E. Sezgin and K. S. Stelle,
               Nucl.Phys. B456 (1995) 669, hep-th/9508042

\bibitem{Berg}  E. Bergshoeff, {\it p-Branes, D-Branes and M-Branes},
               hep-th/9611099

\bibitem{MS1} T. Maki and K. Shiraishi, 
              Class. Quant. Grav, 10 (1993) 2171;
              Prog. Theor. Phys. 90 (1993)  1259

\bibitem{LMPX} H. Lu, S. Mukherji, C. N. Pope and K.-W. Xu,
               Phys.Rev. D55 (1997) 7926, hep-th/9610107

\bibitem{Singh} H. Singh, Phys.Lett. B419 (1998) 195, hep-th/9710189


\bibitem{JMO} B. Janssen, P. Meessen and T. Ort{\'\i}n,
              Phys.Lett. B453 (1999) 229, hep-th/9901078

\end{thebibliography}
\end{document}